# On the Principle of Accountability: Challenges for Smart Homes & Cybersecurity[1]

*Lachlan Urquhart* [2] *and Jiahong Chen*[3]

ABSTRACT: This chapter introduces the 'Accountability Principle' and its role in data protection (DP) governance. We focus on what accountability means in the context of cybersecurity management in smart homes, considering the EU General Data Protection Law (GDPR) requirements to secure personal data. This discussion sits against the backdrop of two key new developments in data protection law. Firstly, the law is moving into the home, due to narrowing of the so called 'household exemption'. Concurrently, household occupants may now have legal responsibilities to comply with the GDPR, as they find themselves jointly responsible for compliance, as they are possibly held to determine the means and purposes of data collection with IoT device vendors. As a complex socio-technical space, we consider the interactions between accountability requirements and the competencies of this new class of 'domestic data controllers' (DDC). Specifically, we consider the value and limitations of edge-based security analytics to manage smart home cybersecurity risks, reviewing a range of prototypes and studies of their use. We also reflect on interpersonal power dynamics in the domestic setting e.g. device control; existing social practices around privacy and security management in smart homes; and usability issues that may hamper DDCs ability to rely on such solutions. We conclude by reflecting on 1) the need for collective security management in homes and 2) the increasingly complex divisions of responsibility in smart homes between device users, account holders, IoT device/software/firmware vendors, and third parties.

## 2.1 Introduction

Consider the following scenarios. A teenage son wants to lodge a *right to be forgotten request* with his father for compromising smart camera footage at their family home?[4] Or a student vacating her flat submits a *data portability* request to her flatmate to obtain the heating schedule built up in their learning thermostat?[5] A grandparent living with their adult children is concerned about them monitoring his diet and submits a *subject access request* to their daughter to see what the smart fridge knows? These situations may seem far-fetched, but this paper unpacks how data protection obligations between household members are increasingly being recalibrated, making these scenarios more likely in the future. This is because the law is moving into the home, due to narrowing of the so called 'household exemption'. Concurrently, household occupants may now have legal responsibilities to comply with the GDPR, as they find themselves jointly responsible for compliance, as they are possibly held determine the means and purposes of data collection with IoT device vendors.

---

[1] This work draws on and extends on the following published/in press papers: Urquhart *et al*. [1], Chen *et al*. [2], Urquhart and Chen [3]. This work was supported by the Engineering and Physical Sciences Research Council [grant numbers EP/R03351X/1, EP/M02315X/1, EP/T022493/1]. This paper is a preprint of a chapter for an upcoming IET book (Crabtree, A. Mortier, R. Haddadi, H. (Eds) *Privacy by Design for the Internet of Things: Building Accountability and Security*). The copy of record will be available at the IET Digital Library.
[2] Lecturer in Technology Law, University of Edinburgh. Visiting Researcher, Horizon Digital Economy Research Institute, University of Nottingham. Lachlan.urquhart@ed.ac.uk
[3] Research Fellow in Information Technology Law, Horizon Digital Economy Research Institute, University of Nottingham. Jiahong.chen@nottingham.ac.uk
[4] A recent Dutch court ruling, albeit in a social media context, has indicated such a possibility. See [4].

To focus our discussion, in this chapter, we examine one set of near-future data protection problems: what the EU General Data Protection Regulation (GDPR) obligations around the *accountability principle* will require in smart homes for *cybersecurity*. This involves unpacking what cybersecurity duties household occupants may have and how they should manage these.

This chapter is composed of 3 primary sections.

We will begin in the introduction by setting some brief context particularly around the high-level message which is: accountability is an important element of the GDPR but changes in data protection law practice mean individuals operating IoT devices in smart homes may need to find ways of providing demonstrations of accountability to data subjects i.e. their family and friends. To unpack this, we focus on demonstrations in relation to personal data information security requirements in the GPDR.

In Section 2.2 we provide an overview of what the accountability principle is and what it might require to be realised in practice. To do this, we firstly consider how it has emerged historically as a governance tool, before turning to how it is framed in Article 5(2) of the EU General Data Protection Regulation. We reflect on the obligations it creates for data controllers and what they need to do in order to satisfy the principle. In this section we will focus on 3 elements. Firstly, we situate accountability in the context of the GDPR more widely, considering it in conjunction with Article 24 GDPR (which unpacks the wider responsibilities of data controllers). We consider the implications of a broad reading of accountability, particularly in relation to compliance strategies for the GDPR and difficulties of prioritising different elements of the legislation. We offer a set of requirements to support this. Secondly, we unpack who accountability is owed to and why this is important, considering accountability requirements that domestic data controllers need to attend to (this theme is explored in more depth in Section 2.4). Thirdly, we reflect on what form a 'demonstration of accountability' might take (e.g. record keeping, privacy impact assessment, privacy by design etc.). Again, this is to raise opportunities controllers have and to explore what some of these might look like. We conclude this section by reflecting on the role of technical measures in demonstrating accountability for compliance with GDPR security requirements in Article 32 and Article 5(1)(f).

We then turn in Section 2.3 to unpacking why we argue domestic data controllers exist as a class of controllers, and how we have reached this position legally. In this regard we reflect on the two aforementioned trends occurring in data protection law around the narrowing of the household exemption and joint controllership broadening.

---

[5] In practice this would not work, as portability does not work this way (it does not cover inferences, i.e. the heating schedule, but just the 'raw' personal data).

We will document what each concept means in data protection law in the light of recent case law which underpins these shifts. We will raise some of the implications here, particularly stressing that it leads to DDCs having responsibilities in relation to demonstrations of accountability, which we pick up on in more detail in the rest of this section.

In Section 2.4, we adopt a more exploratory perspective, aligning the first two parts to pose the question: how can domestic data controllers in smart home demonstrate accountability to data subjects in the home? We will focus on demonstrations in relation to security obligations in the GDPR. As Article 32 talks of both *organisational and technical* measures, we will use this as an opportunity to think about demonstrations under these 2 broad headings, with particular attention to technical measures (but how they work within the organisation of the home). We reflect on the nature of the home as a setting for demonstrating accountability, consider the promise of smart home security technologies but also their limitations. We consider how DP law might impact the domestic social order, particularly with interpersonal relationships and concerns of control, access, permissions and power. Despite these new responsibilities, DDCs are likely to still be domestic users of technology who just happen to be in a position of authority due to hierarchies in the home or from the technology. Thus, we will conclude by questioning the differentiated responsibilities between vendors and DDCs[6] to manage accountability responsibilities.

## 2.2 The Principle of Accountability

In order to understand the accountability principle, we will briefly consider how it has emerged as a data protection tool before turning its current instantiation in Article 5(2) GDPR.

### *2.2.1 Trajectory from the OECD 1980 to GDPR 2016.*

Despite only explicitly appearing once in the EU GDPR [1], the accountability principle has been key in data protection (DP) policy for decades, particularly as a means of enforcing fundamental principles of DP law [6]. Whilst not tightly defined in formulating requirements for action, its flexibility to can be a strength, particularly as it enables more innovative technological and organisational approaches for data protection compliance.

The principle has its roots in early OECD data protection guidelines from 1980 [7]. These were recently updated in 2013 and simply state 'A data controller should be accountable for complying with measures which give effect to the principles stated above'"[7]. The principles referred to include 'collection limitation; data quality; purpose specification; use limitation; security safeguards; openness and individual participation'.[8] Apart from the last two[9], these are broadly similar to the principles that existed in Article

---

[6] Concept of DDCs in Flintham *et al*. [5].
[7] Para 14.
[8] Para 14.
[9] Paras 12 and 13.

6 of the former EU Data Protection Directive 1995 and Article 5(1) of the current EU General Data Protection Regulation (GDPR) 2016.

We return to EU law below, but for now we want to briefly pick up on these final two OECD principles. They state accountability would also require measures that involve greater 'openness about developments, practice and policies with respect to personal data' and also measures to enable rights of individuals in relation to their data.[10] This would require measures for individuals to: obtain confirmation that a controller has their data; to have such data communicated within a reasonable time where that can be for a non-excessive charge, in a reasonable manner and in an intelligible form; to obtain some justification is a request is rejected; and to challenge the data, including having it erased, rectified, completed or amended.

The latter two principles broaden the scope of accountability, covering similar ground to the data subject rights around access, restriction, notification, erasure, portability, and objection in Articles 15-21 GDPR. Openness broadly translates to the transparency and information provision sections in Articles 12-14 GDPR. As we discuss, strict reading of Article 5(2) GDPR relates only to compliance with Article 5(1) hence the OECD framing is broader in content than the EU. However, as we argue below, Article 5(1) needs to be read in conjunction with other GDPR provisions, such as Article 24 on broader data controller responsibilities. This in turn makes it wider (arguably requiring compliance and demonstrations of this with the entire GDPR).[11] However, a key difference between the GDPR and OECD framings of accountability is the latter does not mandate a 'demonstration' of accountability, a point we return to below.

A number of pre-GDPR endeavours have sought to clarify the scope of what accountability means. The Galway [8] and Paris [9] Projects culminating in the Madrid Resolution [10] are three high profile multi-stakeholder attempts[12]. We draw on details from these at different times below, where they help us to understand what a demonstration of accountability is (Paris and Galway Projects) or who it is owed to (Madrid Resolution).

### 2.2.2 *Article 5(2) GDPR and the obligations it creates for data controllers*

The deceptively simple Article 5(2) merely states that 'The controller shall be responsible for, and be able to demonstrate compliance with, paragraph 1'. The principles in paragraph 1 relate to lawful processing, purpose limitation, data minimisation, storage limitation, accuracy, and integrity & confidentiality principles. Explicit inclusion of the accountability principle in the GDPR is new, where it was only implied in the Data Protection Directive 1995 [11]. Despite its short length, it raises a lot of responsibilities for controllers. Firstly, it establishes a *substantive* responsibility for controllers to comply with the series of data protection principles in Articles 5(1)(a)-(f). Secondly, it creates a *procedural* requirement for controllers to find ways to demonstrate their compliance with these principles [1]. The importance of accountability is thus clear in the sense that it severs as a meta-principle that defines how other principles should be observed. Despite this clarity, questions quickly begin to emerge. Is it only these compliance with these principles? what does a demonstration of accountability look like? Is this defined in the law? Who is it owed to? Are there different requirements from a demonstration depending whom it is directed to?

---

[10] Para 12.
[11] See discussion in Urquhart *et al*. [1] on this point.
[12] Business, government, academia.

Furthermore, Urquhart *et al.* [1] argue that Article 5(2) needs to be read in light of wider responsibilities of data controllers detailed in Article 24 GDPR. When this is done, the scope of the provision is much wider, arguably requiring demonstrations of accountability for the entire GDPR. The text of Article 24 states:

> the nature, scope, context and purposes of processing as well as the risks of varying likelihood and severity for the rights and freedoms of natural persons, the controller shall *implement appropriate technical and organisational measures* to ensure and to be able to *demonstrate* that processing is performed in accordance with this Regulation. (Article 24(1) GDPR 2016, emphasis added).

There are many elements to unpack in Article 24, from questions of processing to risks to what measures are necessary and again, what a demonstration might necessitate. However, given Article 24 goes far beyond just the Article 5(1) principles, it could be quite overwhelming for controllers to determine which elements of the GDPR should be prioritised (especially as there is no real hierarchy within the law of what to prioritise). Conscious of this, Urquhart *et al.* [1] attempt to break down and cluster the responsibilities of controllers into a series of 7 Accountability Requirements, as seen in Table 2.1 below (which is discussed in more detail in relation to requirement 6). From a procedural and substantive perspective, a data protection impact assessment (Article 40) can be a useful tool for both surfacing processing risks whilst also providing a physical document that can be a demonstration. The similarities between Articles 24, 25 (on data protection by design and default) and 32 (on managing data security) are also significant, where turn to technology safeguards is a key policy direction. Technologists are being drawn into the regulatory fray and system architecture is an acknowledged route to address wider responsibilities of data controllers when processing personal data. As this is arguably to comply with the entirety of GDPR and demonstrate how this is done, the scope for convergence of solutions that address Articles 24, 25 and 32 here is sizable i.e. deploying information privacy preserving architectures by default could also satisfy demonstration of controller responsibilities and good security practices.

### 2.2.3 *Accountable to whom?*

The GDPR does not state who accountability is owed to or what it needs to involve, which the former Article 29 Working Party have argued is by design, to enable flexibility of application on a case by case basis [12]. This mirrors the OECD position where their Guidelines 'do not prescribe to whom the controller should be accountable (the 'accountee'), nor what this relationship should look like' [13]. The GDPR does not constrain this either. In practice it is useful to ground who should be targeted and the Madrid Resolution stated it should be, *at least the data subject and the data protection regulator*. However, in smart homes where DDCs may owe accountability to these parties, whilst themselves being users, it is less clear how this might manifest. For organisations, as we see below, they have tools at their disposal to demonstrate accountability. How these translate to the domestic, complex socio-technical context of the home, is less clear. This is to state the accountability requirements that domestic data controllers might need to attend to (picking up on this theme in more depth in Section 2.4). How might they demonstrate accountability to a data subject? Much less a data protection authority? Who else might have a vested interest? IoT technology vendors?

### 2.2.4  What form demonstrations of accountability might take?

As noted at the beginning, accountability is a flexible notion. The fact that what a demonstration requires is not prescribed in law means creative approaches can emerge (e.g. as Urquhart *et al.* [1] argue, Databox can be a demonstration). Nevertheless, in practice we need to pin down what a demonstrable account requires to operationalise it to consider the range of options available to controllers [1]. Raab [6] has argued, how an account is framed can differ in strength from a basic level of documenting what has been done by a controller, to enabling questioning or subjects contesting the story, leading up to sanctions. after consulting guidance from the UK ICO [11], EDPS [14], and the pre GDPR Galway [8] and Paris [9] projects to suggest what demonstrations of compliance, could require, clustering guidance into technical and organizational forms. They are:

> **Technical measures**
> Data protection by design and default; including use of anonymization, pseudonymization and end-to-end encryption; IT security risk management.

> **Organisational measures**
> Assigning DP officers (DPOs); prior consultations; certification schemes; DPIAs; transparent policies; documentation and record keeping on processing for organizations with over 250 staff; internal compliance and audits for effectiveness of approaches; training [1].

As the guidance on accountability is geared towards organisations, the focus is often framed as developing, as the EDPS [14] puts it, a 'culture of accountability' and not just a bureaucratic 'box ticking exercise' [15]. In the Galway [8] and Paris [9] projects they unpack what a culture or demonstrable accountability might look like, ranging from internal governance structures for organisational compliance with DP standards, to enforcement bodies, training on privacy, leadership and risk analysis.

How might this translate in a home environment where the controller is another household member, and data subjects are children, spouses, extended family and friends, trades people? We consider both in part III, with particular focus on examples and uses of edge computing-based security management. For now, we want to conclude this section by considering what the GDPR precisely states in relation to cybersecurity accountability obligations, to pinpoint what domestic data controllers need to substantively and procedurally address.

*Table 2.1    Accountability requirements in GDPR (from Urquhart et al. [1])*

| Accountability requirement | Source in GDPR |
| --- | --- |
| 1. Limiting initial data collection | Purpose limitation Article 5(1b); data minimization Article 5(1c); storage limitation Article 5(1e) |
| 2. Restrictions on international data transfer | Data sent outside Europe on basis of adequacy decision Articles 44 and 45; binding corporate rules Article 47; appropriate safeguards Article 46 |
| 3. Responding to the spectrum of control rights | Right to access Article 15; to rectification Article 16; to object Article 21; to restrict Article 18; to portability Article 20; to erasure Article 17; information supply chain (passing down requests for rectification, erasure, restriction) Article 19 |
| 4. Guaranteeing greater transparency rights | Transparency of information Article 12; rights to provision of information Articles 13 and 14; algorithmic profiling Article 22; record keeping Article 30 |
| 5. Ensuring lawfulness of processing | Legality based on specific grounds (Article 5(1a) and Article 6, e.g. performance of contract legitimate interest); consent requirements Article 4 (11), Article 7, Article 8 and Article 9 |
| 6. Protecting data storage and security | Accuracy of data Article 5(1d); integrity and confidentiality Article 5(1f); breach notification to authorities Article 33 and to data subject Article 34; security of processing Article 32 |
| 7. Articulating and responding to processing responsibilities | Articulating responsibilities: Data Protection Impact Assessments Article 35; certifications including seals, marks and certification bodies Articles 42 and 43; new codes of conduct Articles 40 and 41 Responding to responsibilities: DPO Articles 37 and 39; DPbD Article 25 |

Requirement 6 states in Table 2.1 states the key dimensions of accountability for security are to 'protect data storage and security', based on 'Accuracy of data Article 5(1d); integrity and confidentiality Article 5(1f); breach notification to authorities Article 33 and to data subject Article 34; security of processing Article 32'.

Articles 33 and 34 are around breach notification, notably dependent on the nature of breaches (scale, data compromised) and 72-hour time frames for doing this. Article 5(1)(d) requires that data be accurate and up to date whilst inaccurate data be also handled swiftly ('the accuracy principle'). However, in this paper we focus on the technical or organisational aspects of the following two GDPR provisions:

- Article 5 (1)(f) requires 'appropriate security' with processing to guard against 'unauthorised or unlawful processing and against accidental loss, destruction or damage' with 'appropriate *technical or organisational* measures'.
- Article 32 requires *technical and organisational* safeguards for security to be built into processing, proportionate to risks and considering the state of the art, costs, and nature of processing.

When viewed alongside Articles 24 and 25, we see an increased focus on technical and organisational measures in the GDPR. For security, this could include controllers taking technical steps such as use of anonymisation; pseudonymisation; end-to-end encryption;[13] regular testing of safeguard measures; mechanisms to ensure ongoing 'confidentiality, integrity, availability and resilience' of systems; provisions giving the ability to restore access to data quickly.[14]

As we will see in Section 2.3, there is a growing role for home occupants in managing DP compliance, and thus data security. The emergence of smart home security tools could be a mechanism that satisfies elements mentioned above. However, we want to explore the organisational dimensions of domestic compliance alongside the technical measures that might assist DDCs in demonstrating accountability, as the two cannot be separated easily. Furthermore, we focus on demonstrations to data subjects, as opposed to authorities in this paper, given the complex social issues this raises. We return to these issues of these systems, and organisational security management in smart homes in depth in Section 2.4, but firstly need to unpack why we are even talking about DDCs having a role to play here. Thus, in the next section we explain why DDCs exist as a class of controllers, and how we have reached this position legally. This involves two fundamental shifts in EU case law: a broadening of the notion of a data controller and a narrowing of the household exemption in DP law.

## 2.3 Data Protection in the Home?

To the extent that the principle of accountability imposes the major DP compliance duties on the data controller, it is of paramount practical importance to first identify the controller – or controllers – and then deicide the scope of their responsibilities. Ascertaining who is the *de facto* data controller, however, is not always straightforward, and the increasing prevalence of IoT devices used in spaces of various nature has brought in even more legal uncertainties, not least for the wider range of actors involved and heavier reliance on technical protocols. As will be elaborated below, the GDPR has set out several rules in order for the assignment of accountability duties to reflect the nature, risks and expectations regarding certain types of data processing activities [2]. Most importantly, the household exemption effectively creates a carve-out of the scope of the GDPR whereby domestic data controllers (DDCs) may be exempt from demonstrating compliance with the DP principles; joint controllership, on the other hand, establishes collective accountability for data controllers who jointly exercise control over the use of data.

Both the household exemption and joint controllership are therefore important considerations for one to fulfil their duties under the principle of accountability. These two legal notions, however, have been subject to judicial and scholarly debates since the time of the Data Protection Directive. Recent case law and the entry into force of the GDPR have added further dimensions to the complex application to a domestic IoT setting. By reviewing the developments of these two concepts as well as their interactions, we aim to unpack their implications for DDCs in relation to their observance of the accountability principle.

---

[13] Article 25 suggests these first few too.
[14] Article 32(1).

### 2.3.1 *The household exemption and its (in)applicability to domestic IoT*

Under the exemption provided for by Article 2(2), the GDPR 'does not apply to the processing of personal data […] by a natural person in the course of a purely personal or household activity'. Any qualified domestic use of personal data, accordingly, would not be subject to any DP principles or restrictions, including the accountability principle. Two conditions of this provision immediately stand out: 'by a natural person' and 'purely personal or household activity'.

The implication of the first condition is straightforward: This exemption applies only to individuals and not organisations, regardless of the possible household nature. Manufacturers of IoT devices or providers of IoT services are thus simply unable to claim this exemption even if the use of personal data proves indeed for purely personal purposes. For end-users of smart home technologies, there is room for a claim that DP law does not apply to them as long as the second condition is also met. This, in practice, would mean that such users, even if they would otherwise qualify as a data controller, do not need to comply, or demonstrate compliance, with any of the DP requirements. A secondary question would then arise as to whether other organisational joint controllers, if any, should be expected to demonstrate the end-user's compliance as part of their accountability duties. We will discuss that issue in further detail in Section 2.4. Here, from the perspective of the users, the connection between the household exemption and the accountability principle is rather obvious: The application of former would lead to the complete exclusion of the latter.

The much more complicated prong of the household exemption lies in the second condition, namely the purely personal nature of the processing in question. The recitals of GDPR (the non-binding statements in the preamble) as well as regulatory bodies have provided some general clarifications, but it is through the jurisprudence developed in the cases decided by the CJEU that specific complexities exhibit themselves in particular scenarios. In a highly relevant case, *Ryneš* [16], the Court was asked whether the operation of a CCTV camera on a residential building may be considered as a 'purely personal or household activity'. The Court rejected the claim on the basis that the 'video surveillance […] covers, […] partially, a public space and is accordingly directed outwards from the private setting of the person processing the data in that manner' (para 33). This interpretation was followed in a later case *Asociaţia de Proprietari bloc M5A-ScaraA* [17], in which the Court decided that the installation and operation of CCTV camera filming 'the common parts of [an apartment] building in co-ownership and the approach to it' (para 49) are also subject to the GDPR.

It should be noted that in either case, the domestic *purpose* or *intention* was not questioned by the Court. In fact, the Court was fully aware that the data controller – i.e. the operator of the CCTV system – may have legitimate interests in 'the protection of the property, health and life of his family and himself' [16] (para 34) or 'ensuring the safety and protection of individuals and property' [17] (para 33). It is the operative *method* of the CCTV system that came to the central point of the Court's analysis. The fact that the such a surveillance system may casually capture and store the image of individuals outside the data controller's family in an electronic format will suffice for such activities to fall outside the scope of the household exemption. The implication of this judgment for smart home owners can be profound to the extent that many IoT devices are capable, or indeed designed, to collect data from a space potential beyond the physical boundaries of one's home. Visitors, neighbours, or even passers-by may be affected by, for example, the accidental collection of data from their smartphones when they approach the sensory remit of the user's smart home system.

Also, worth pointing out is that the CJEU has never ruled in favour of a claim of the household exemption. Rather, the Court has consistently taken a highly restrictive approach to interpreting this provision. By ruling out the applicability of the exemption to various cases, the scope of this exemption is increasingly shrinking, which may have an impact on any IoT end-users hoping to benefit from this exemption. The rationale of this interpretative approach is perhaps not difficult to understand: As discussed above, triggering the household exemption would mean the GDPR ceases to apply altogether to the case concerned, which would potentially create a regulatory vacuum to the detriment of the data subject. While the Court has laid down clear rules on what would *not* count as purely personal - data accessible by an unrestricted number of people, or concerns a public space beyond the private sphere [18] (para 42) – the only cases unquestionably exempt are those explicitly provided for in Recital 18 GDPR, namely 'correspondence and the holding of addresses, or social networking and online activity undertaken within the context of such activities'.

As such, the circumstances under which a claim of the household exemption can be reasonably made by an end-user will need to be examined on a case-by-case basis. If the use of personal data in these context does not pass the threshold set out in the case-law, the user would have to demonstrate compliance with the accountability principle. While the Court has never directly addressed Article 5(2) GDPR (or Article 6(2) DPD), it seems obvious that accountability suggests some form of duty of care to be assumed by the data controller even concerning merely household activities, which would entail an assessment of the technical options in the light of the interests and risks of the parties involved. As AG Jääskinen opined in *Google Spain and Google* [19], 'Article 6(2) of the Directive obliges [data controllers] to weigh the interests of the data controller, or third parties in whose interest the processing is exercised, against those of the data subject.' (para 107) In *M5A-ScaraA* [17], the Court also stated that 'the proportionality of the data processing by a video surveillance device must be assessed by taking into account the specific methods of installing and operating that device' (para 50) and 'the controller must examine, for example, whether it is sufficient that the video surveillance operates only at night or outside normal working hours, and block or obscure the images taken in areas where surveillance is unnecessary.' (para 51) How such considerations can be demonstrable by domestic users in practice, however, is a different and yet even more challenging issue.

### 2.3.2    *Domestic and non-domestic joint controllership in domestic IoT*

Another concept crucial to implementing the accountability requirements is joint controllership, as it defines how DP obligations are shared among a group of controllers who make co-decisions on how personal data are processed. Article 26(1) provides that 'Where two or more controllers jointly determine the purposes and means of processing, they shall be joint controllers.' The same provision goes on to require joint controllers to 'in a transparent manner determine their respective responsibilities for compliance with the obligations under this Regulation'. This definition seems to suggest that joint controllership arises from a unanimous decision by all controllers with regard to how they DP duties are split and discharged by each of them. However, the allocation of duties cannot be determined on an arbitrary or unreasonable basis. Instead, such an arrangement must 'duly reflect the respective roles and relationships of the joint controllers *vis-à-vis* the data subjects.' This has also been emphasised in the opinions issued by the Article 29 Working Party. The Working Party has specifically highlighted the importance to establish 'clear and equally effective allocation of obligations and responsibilities' [20] that genuinely reflects the legal relationship of between the joint controllers. For example, if a contract substantially gives one an entity material powers to decide how data is processed, but formally only assigns a different entity with less influence as a sole controller, such an assignment would be invalid and the former entity would remain liable as a joint controller.

Apart from the scenarios involving contractual arrangements, which are clearly covered by the definition of joint controllership, the CJEU has decided on a number of more complicated cases where such a contractual relationship does not clearly exist. For example, in *Jehovan todistajat* [18], the Court decided that the Jehovah's Witnesses Community was a joint controller with its members with regard to the collection and use of personal data through door-to-door preaching. The judgment was concluded on the basis that the 'preaching activity is […] organised, coordinated and encouraged by that community' (para 70), regardless of the lack of formal instructions issued, or actual access to the data, by the Community (paras 67, 69). The mere organisational structure facilitating the processing of personal data was considered sufficient to give rise to joint controllership.

Joint controllership may also result from activities aligned by technical protocols. The Court considered in *Fashion ID* [21] whether the operator of a website became a joint controller with Facebook by placing a 'Like' button on a webpage, which would trigger and enable Facebook to collect personal data from a visitor. This does not depend on any communications or existing legal relationship between the website and Facebook, and as such, the formation of joint controllership can be purely a matter of technical settings. Although *Fashion ID* does not concern a smart home context, the implications for domestic IoT users can be profound as the two scenarios bear some resemblances: One party controls the switch of a system, whose architecture is designed by the other party to allow the latter to collect personal data from third parties. In Chen *et al.* [2]'s words, the former possesses *operational control* and the latter *schematic control*. Data collection would not take place without the action of either party, but their cooperation does not rely on a formal mutual agreement, but simply by ways of technical configurations.

The expanding coverage of joint controllership by the CJEU means that the 'co-decision' by joint controllers does not have to take the form of contractual arrangements, but can be achieved simply with a much looser relation enabled by organisational or technical configurations. Recognising this point is especially crucial for compliance with the accountability principle because data controllers are under a clear obligation to 'implement appropriate technical and organisational measures' (Article 24 GDPR). The nature of the joint controllership can be a critical first step in establishing what measures should be put in place to coordinate responsibilities and document compliance. In fact, the Court has noted that the joint controllers 'may be involved at different stages of that processing of personal data and to different degrees, so that the level of responsibility of each of them must be assessed with regard to all the relevant circumstances of the particular case' [22] (para 43). For example, in *Fashion ID*, the Court unequivocally pointed out that the duties to obtain consent from, and to provide information to, visitors of a website with an embedded Facebook 'Like' button fall on the operator of the website, not Facebook, even though the data are transmitted only to the latter [21] (paras 98-106). A comparison can be made with the case of domestic use of IoT technologies, where the homeowner, if held jointly responsible with an IoT service provider who collects data potentially from non-members of the family, may be under a duty to ensure there is a legitimising basis and to inform the affected parties (e.g. guests) of the details.

### 2.3.3 *Accountability shared between domestic and non-domestic controllers*

Having discussed the implications of the narrowing scope of the household exemption and the widening application of joint controllership in case-law, now let us consider how these complexities would play out in observing the accountability principle in a variety of circumstances in an IoT-enabled smart home. Table 2.2 below shows different combinations of the nature of the processing and the category of the controllers. It should be noted that a domestic data controller (DDC) here refers simply to an individual processing personal data with an IoT device in their home, and does not necessarily imply such uses are for domestic purposes only, who may or may not be covered by the household exemption. Likewise, a non-domestic data controller (non-DDC) is defined here as a controller other than a DDC, who does not have to process the data for a professional or commercial purpose.

*Table 2.2   Data controller(s) responsible to demonstrate accountability*

|  | **DDC only** | **DDC & non-DDC** | **non-DDC only** |
|---|---|---|---|
| **Purely domestic use** | Nobody | non-DDC * | N/A |
| **Non-domestic use** | DDC † | DDC & non-DDC ‡ | non-DDC § |

It is noteworthy that even if a particular use of personal data is within the course of a 'purely personal or household activity', the household exemption would not apply to the non-DDC, meaning that the latter would remain responsible for the processing in question. This is because the household exemption operates essentially on a controller-specific basis. Recital 18 GDPR makes it clear that 'this Regulation applies to controllers or processors which provide the means for processing personal data for […] personal or household activities.' As a result, the non-DDC may be held responsible as the sole controller in two scenarios, one where the domestic user does exercise control on the processing but is exempt from the duties due to the domestic nature of the processing (scenario *), and the other where the domestic user has no control at all (scenario §). While it is clear in the latter case that the non-DDC must demonstrate accountability concerning the full range of DP obligations, this is open to question in the former case where certain duties would have fallen on the DDC had the household exemption not applied. For example, consider the situation where a homeowner uses a smart thermostat in a guest room occupied by a visiting family guest, which sends room temperature data to the vendor's server that are accessible only to the homeowner. Assuming such data are personal data and this is a purely household use of such data, would the vendor, as a non-DDC, be placed under any primary or secondary obligations to demonstrate that, say, there is a legitimate basis on the homeowner's part to use such data, or adequate information has been given to the guest? One interpretation of the household exemption can be that these duties would not apply at all once the exemption has been established, meaning that no demonstration of accountability would be required when it comes to such duties. However, it can be an equally valid contention that such duties do not apply, but just *to the DDC*, and the exemption does not preclude the non-DDC's duties on these matters if they have not already been fulfilled by the domestic controller. Such uncertainties would call for further regulatory or judicial guidance.

Another challenge is where it is incumbent on the DDC to demonstrate compliance with the DP principles, whether solely (scenario †) or jointly with the professional controller (scenario ‡), how their accountability can be demonstrated and assessed. The subtle differences mirrored in varying interpretations of the tensions between DDCs, non-DDCs and data subjects may remarkably affect our understanding of what is 'accountable' and what is 'demonstrable' in a domestic IoT environment. Such nuances are not merely a legal fiction but rather a relational reflection of the complex socio-technical attributes embedded in today's smart homes. As will be seen in the next section, empirical evidence has suggested this is indeed a highly sophisticated landscape. Effective policymaking as well as enforcement of the accountability principle would therefore depend on a meticulous capture of the role of various solutions in both technical and organisational terms in managing accountability. It is in this regard that we now turn to the HCI scholarship for further answers.

## 2.4 Accountable domestic data controllers

In this part we align discussions from the two previous sections to pose questions around how domestic data controllers in smart homes can demonstrate accountability to data subjects in the home. As an emerging domain, we take this opportunity to explore related literature, and pose some questions to be addressed to understand what DP law coming into the home may mean in practical terms. We are interested in both the form and challenges that may arise in creating *demonstrations of accountability*. DDCs may have to provide appropriate security (Article 5(2) GDPR) and use technical and organisational measures to support secure data management (Article 32 GPDR). What this will look like in the domestic context requires unpacking, particularly around interpersonal relationships in homes.

We focus here on technical measures, but consider how the organisational deployment setting (i.e. the home) will shape how technical measures might work or be adopted. Below, we turn to contemporary empirical and technical work from fields of human computer interaction and usable privacy/security, to help us unpack the following lines of inquiry:

- What technical measures exist for supporting demonstrations of accountability, specifically focusing on smart home security management tools?
- What occupant privacy and security management practices exist in smart homes, particularly around interpersonal relationships?
- What socio-technical and usability problems might these tools raise with DDCs using them?
- How might these discussions intersect with GDPR compliance requirements for DDCs, particularly for demonstrations of accountability?

A fundamental tension raised by the shifts outlined above is that giving DP responsibilities to DDCs is *they lack both resources and skills of organisations* that the legal framework was designed to apply to. Thus, there can be shared responsibility with vendors of smart home technologies or even those providing smart home security management tech (depending on the architecture). *Ordinary* organizational mechanisms are unlikely to work here e.g. will a partner be auditing the quality of their spouse's password? Will they conduct a review of breach management strategies? What might domestic processing record keeping look like here and the description of measures taken?[15]

Another is the compliance *setting*. It is not a company with organisational resources for record keeping capabilities or capacity to know when to do a data protection impact assessment or hire a data protection officer (many normal steps seen for accountability). Instead, it is a home; space where cohabitants live their daily lives. Thus, translating regulatory concepts to this domain could feel forced. Should regulatory norms structure domestic life? Domesticating technology itself can take time [23] and designing technologies so they can be embedded in daily lives requires an appreciation of what that life looks like [24]. The same will be true for *designing technical and organisational measures for demonstrating accountability*. We need to understand how security is managed currently in smart homes, and thus we firstly consider what tools there are and the context they will need to operate in i.e. what do occupants currently do?

### 2.4.1 Smart home cybersecurity management tools

Whilst traditionally (and still predominantly) the system architectures of IoT devices are centralised and cloud based, we are seeing increased use of distributed analytics and storage at the edge of networks. In terms of data handling responsibilities, managing these systems may increase the role for users and others (e.g. DDCs) at the edge of the network. This is often by design, in the case of some edge-based storage systems, such as personal information management systems (PIMS) like *Databox, HAT, MyDex* etc.[16] It is recognised in privacy policymaking and privacy engineering communities that PIMS can help users determine how their data is used, and increase control over who has access to it [26, 27]. Similarly, distributed data analytics can enable new privacy preserving forms of federated machine learning which can address ethical and governance concerns around data harvesting, severing personal data from users and associated privacy harms stemming from big data analytics. This shift to 'small data' led systems [28, 29] can empower users, address data protection compliance concerns, particularly around opacity of data flows and power asymmetries around access by third parties [1]. Nevertheless, these systems still require (sometimes high levels of) oversight and awareness around how to do this. Given varying skillsets and motivations to manage privacy and security, this could be problematic. Interface design often does not help either (as we return to in topic 3).

These systems open up scope for edge-based systems for security management to enabling smarter network monitoring and security management tools. There are a range of commercial and research led offerings, including *Cujo* [30], *IoT Keeper* [31] and *Sense*

---

[15] Article 30 GDPR – documenting details about processing e.g. purposes of processing, categories of data and subjects etc but importantly, under Article 30(1)(g): 'where possible, a general description of the technical and organisational security measures referred to in Article 32(1).'
[16] See Urquhart *et al*. [25] for more details.

[32]. We describe a few examples below, including both prototypes and commercial products.

- *IoT Inspector* examines device firmware for security vulnerabilities after it is uploaded to the platform, analysed and risks reported back [33, 34]. It is geared towards a range of industry stakeholders, particularly vendors, and thus could be useful for joint controllers, where responsibility lies with them, in addition to DDCs.[17]
- *Fingbox* is a consumer orientated physical box that can be bought for the home. It monitors network activity (including open ports), conducts vulnerability analysis, blocks intruders or unknown devices, and observes who is on the network.[18]
- *Wireshark* is an open source, freely available network protocol analyser that provides granular analysis on network activity by sniffing packets. It can be used for intrusion detection too [35]. For an everyday user, there may be skillset requirements in interpreting this data.
- *Homesnitch* is a prototype that seeks to increase transparency and control over domestic networks by classifying device behaviour. As opposed to just reporting packet flow it 'learns' about behaviour in order to report on what this might mean e.g. 'downloading firmware, receiving a configuration change, and sending video to a remote user' [36]. This could be more contextually useful for users.
- *IoT Sentinel* is another prototype architecture where the system has a more active role in managing network security. It identifies types of devices on a network, spots any security vulnerabilities and then enforces network rules against these devices automatically. This latter step can range from isolating devices and blocking external access, filtering traffic to prevent data being exfiltrated or notifying users of issues [37].
- The *DADA system* seeks to monitor network traffic for vulnerable devices and unexpected behaviours (e.g. using MUD profiles). It then informs users of different options to deal with vulnerable devices e.g. to block the device from the network etc [38].
- *Aretha* includes an integrated training aspect for end users, firewall management and a network disaggregator [39]. Part of the tool gave users the ability to set directives for the firewall in a user friendly way e.g. with a GUI and in plain language. Despite this, they found there was need to simplify this to enable user adoption, coupled with difficulties in users analysing risk of safety or danger in blocked domains. The researchers posit a greater role for experts pre-formulating black lists that users can then tweak, or using presence of certain metrics to block sources e.g. 'company reputation, jurisdiction, purpose of data collection, retention and disclosure properties'.[19] This tool also offers training, visualisations and a physical presence, all helping strengthen collective user engagement with security

---

[17] Also targeted to corporate users, infrastructure providers, researchers and resellers.
[18] https://www.fing.com/products/fingbox
[19] They highlight that it does a variety of things: 'the probe brought about not just a heightened awareness of privacy concerns—by being a salient and visible physical totem in their living room—but a transformative one, in which privacy management transitioned from being a solo concern anchored with the individual responsible for technology within the home, to one discussed between multiple home stakeholders. The visibility of the probe within the home inspired conversations among household members, and the visualisations provided a common ground'" (p9).

and privacy management.

Whilst these tools show promise as technical measures for DDCs demonstrating accountability to other occupants, we also believe there are complex socio-technical considerations around their possible integration into the home. Thus, we will review some literature around current user security management practices in smart homes.

### 2.4.2   Security (and privacy) management in smart homes

Security does matter to smart home users. A study by Emami-Naeini *et al.* [40] found that whilst not as highly valued as price and features, both privacy and security are important factors when buying new IoT devices. Nevertheless, consumers can find it difficult to obtain information about the security and privacy credentials of a product but do state they may be willing to pay a premium (between 10-30%) for a device with this information.[20]

In the wild, user concerns around security and privacy threats in smart homes vary. Zeng *et al.* [41] find participant concerns focused on physical security e.g. control over smart locks enabling home access. They observed that different levels of technical knowledge shaped their awareness of types of IoT vulnerabilities e.g. skilled users were concerned about HTTPS, less skilled about weak passwords or unsecured Wi-Fi (p 71). They also observed this knowledge shaped how users manage risks from changing behaviour e.g. avoiding speaking in front of Alexa to more skilled participants setting up a separate Wi-Fi networks or blocking traffic.

Looking to smart speakers, as an example smart home system, Huang *et al.* [42] unpack privacy concerns around smart speakers with respect to internal (siblings/flatmates), key for DDCS to consider. These include:

- Concerns about overheard calls,
- unauthorised access to contact details, calendars or reminders on speakers, due to inadequate voice authentication (e.g. it is open to all as opposed to being tied to a specific users' voice).
- speakers being exploited for unintended uses by children or visitors like unauthorised purchases

In terms of managing these concerns, more skilled users might not link any private information with the speaker and they suggest there could be routes for customising speakers to share information users might be comfortable their peers accessing.

As Crabtree *et al.* [43] have noted, within the home, the nature of control and who accesses information is not always framed in 'privacy' terms. Often it is about managing relationships and is part of the work of living with networked systems. As they state, '"privacy" *dissolves* into a heterogeneous array of mundane practices and local concerns that are not primarily to do with the disclosure of personal data, but with managing **who** gets to access what devices, applications, and content'. However, with the emergence of data protection norms within smart homes, it forces reflection on how informational privacy might be reframed, and how those norms structure domestic relationships and practices. We now turn to studies exploring security management in homes in more depth, particularly interpersonal aspects.

---

[20] They argue labelling schemes might offer viable solutions in simplifying information provision.

### 2.4.3   Interpersonal power dynamics in smart homes

Geeng and Roesner [44] provide valuable insights into social *power dynamics* in smart homes, particularly around the 'smart home driver' i.e. the individual who wants to have the device in the home. Drivers have a prominent role when something goes wrong and in mediating access / control over the devices for other occupants. This can differ across the life cycle of devices e.g. when first installing to long term management.[21] They note that drivers are ordinarily men and had more interest in learning about using the devices, thus giving more power and agency in the home, whereas other relied on them to manage change (p 8). Drivers may be more likely to be the DDCs, as the one managing the device and its processing in the home.

Their role can be problematic for other occupants and Zeng *et al.* [41] provide data from those who are not primary but 'incidental' users in the home. They observed incidental users might not have apps installed or even have access to functionality of devices. This ranges from less interest to restrictions from primary users to prevent them from changing settings e.g. restricting change of thermostats. This control can take different forms, from landlords remotely observing transcripts from smart speakers or security camera footage to monitoring use of the home e.g. parties, or even surveilling when other household members leave/arrive at the home (e.g. husband monitoring smart lock). The latter example aligns with earlier studies from Ur *et al.* [45] around impacts on family dynamics from parents observing when children arrive home through smart lock logs. Zeng *et al.* [41] conclude, 'while the people who set up smart homes, particularly early adopters, often treat the technology as a personal hobby, smart homes are fundamentally not personal technologies. As a result, any security and privacy (or other) decisions made by the primary user directly affects other residents and visitor' (p75).

More concerningly, this power could link into Freed *et al.* [46]'s work examining how technology is used for intimate partner violence. They observe how authorised users use functionality that they have access to in order to track, manipulate and threaten their partners. In response, they recommend use of reviews during the design process or managing default settings when systems are adopted (e.g. removing display of recent location information). Clearly in smart homes this could be a risk and the work of Parkin *et al.* [47] responds by developing different usability heuristics to assess risks IoT devices create for intimate partner violence. Recent work by Levy and Schneier [48] further theorises how smart home technologies can facilitate invasion of privacy in intimate relationships by augmenting the coercive power possessed by the member with explicit or implicit authority.

Mundane smart home device *account* management can also raise power issues in families. Goulden [49] explores the way firms, such as Google and Amazon construct hierarchies in the household to coordinate everyday activities. This can involve linking accounts and defining roles in the home e.g. by attributes like age. This can enable control over others accounts and what they can do with and observe uses of devices e.g. remove permissions or delete from family group. Goulden is concerned about the impacts on domestic family life of this type of intervention.

---

[21] e.g. 'Device Selection; Device Installation; Regular Device Use; Things Go Wrong; Long-Term Device Use'.

In smart homes, data is co-constructed and relational, it often does not only identify one individual [50]. This can lead to difficulties for us in thinking about data controllership and subjects. In a home, both DDCs and data subjects may be identifiable e.g. in traces from sensors in the home. Reconciling this issue needs further attention around the duality of how a controller can be both a steward for personal data of others in the home, alongside their own. This is further complicated when we consider the joint nature of controllership with IoT vendors, who may not have the same interests in this data. Family members would likely want to protect the data of other family members (violence and abuse notwithstanding), but, as we know, IoT vendors may seek to monetise and use that data to understand everyday life and make inferences. The power dynamic between controllers and subjects in this domain will remain messy.

Desjardins *et al.* [51] observe different relationships with data in smart homes, and how data is visualised within the home by occupants, developing how data is part of experiencing the home and daily life. They see 'IoT data not as an undefined, ephemeral, position-less, and singular mass (although it might be conceptualized this way): these data are, in fact, of a home, in a home, and part of unique domestic assemblages which are important to recognize and honour when designing for them.' This point further highlights how established labels and categories in DP law around personal data, identifiability and prescriptive responsibilities over data might not translate to the home context, where the messiness of data is part of life. Yet, with DP law entering the home DDCs will face responsibilities for other occupants. They will need to consider how what they do fits with legal terms of art, like personal data, and how they can demonstrate they are protecting data rights of others.

### 2.4.4   *Control struggles in smart homes*

Despite thinking about these near future issues, we need to unpack how control of data and devices is managed *currently* in smart homes, given accountability requires demonstrations from DDCs to subjects (i.e. between family members). Thus, it is valuable to consider what types of control issues smart homes pose for daily life.

Geeng and Roesner [44] consider some of the interpersonal tensions different stakeholders' face in *controlling* domestic smart technologies. Examples include: partners disagreeing about third party access via door lock code (e.g. cleaners); roommates controlling temperature via apps, where others do not use the app; sibling locking each other's media accounts for punishment; parents and children competing for control of an Amazon Echo e.g. with music it plays (p 6). Physical security and safety concerns around co-occupants being able to still use resources despite lack of expertise or account control like lighting, heating and exit/entry. This is particularly important when devices stop working, e.g. in DIY smart homes.

Given the home is a shared space, co-management of devices and how groups manage security of shared resources is an important consideration we return to below. Such relational complexities are perhaps exacerbated by the heterogeneity of governance structure of smart systems, or in Singh *et al.* [52]'s words, the 'systems-of-systems' nature of IoT. Technologies can play a role, they argue, in enabling accountability by supporting control and audit of IoT systems.

Mazurek *et al.* [53] consider how access control across devices is managed in homes, exploring the complexity of creating approaches that work in practice vs how users want them to work. Whilst focused on protecting sensitive data on laptops or phones, it is interesting the lessons about shared device usage for IoT too. Designers need to enable users to create granular access controls, to enable subtle creation of guest accounts (to avoid awkwardness when sharing devices). Interestingly, they noted that in deciding about access, it went beyond *types of files* (work; music; photos etc) and *who* the other user was e.g. partner, parents, friends etc). Instead, factors such as presence and location of both controller and those accessing files was important, as was the device type (e.g. phones more personal) and even time of day (e.g. in the evening where sleep should be occurring). These show what might shape access policies to information and how for IoT.

This is valuable to further reflect on concerns of Goulden [49] that smart homes having formalised models of control, access and roles. Here, we can see, the home is messy, and notions of accountability play out differently there, unlike in institutional security policies with strict hierarchies, safelists and permissions work more effectively. In homes, it needs to be more dynamic and contextual. In terms of how vendors might support accountability, as joint controllers, this is an area for further work on designing more adaptive forms of permissions and account management. One route forward could be from ID management protocols. Rosner and Kenneally [54] have argued in relation to IoT privacy that identity management is important to consider because it involves 'discussion of privacy as control, access management, and selective sharing'. They suggest building systems that prevent linking of identities and unobservability of users by default, in addition to value of protocols such as UMA, which enable user led control of access and selective permissions.[22]

For DDCs, this could be a valuable tool but would require vendors to enable such functionality. In terms of other routes forward, we now consider the importance of engaging others in the home in security management.

### 2.4.5   Towards collective security management in homes?

Geeng and Roesner [44] argue that designers need to be more aware of different relationship types in homes and also consider how to make the account creation process more sensitive to multiple occupancy with shared devices. They suggest IoT designers need to incorporate 'mechanical switches and controls' for basic device functionality like switching on or off. They also suggest need for measures to manage temporary device access for short term occupants and also for deletion/migration of data when people move out. Similarly, Zeng *et al.* [41] recognise systems should 'support multiple distinct user accounts, usability and discoverability of features are critical for secondary, less technical users' e.g. using physical controls and indicators in the home for when being recorded.

Tabassum *et al.* [55] have also looked beyond the walls of the home, to consider how use of smart devices is shared with others outside and the nature of this shared management. Factors such as security are critical for enabling others to access devices to deal, with motivations like helping manage deliveries or emergencies when occupants are away. Enabling remote access to the home with smart locks or safety of pets and older people is another example. However, they noted that this has to be underpinned by trust in the community they are sharing access with and because of this, often it was full access given to external parties, as opposed to more nuanced forms. Here, we can see that DDCs might not just be parties within the home but managing data flows as an external party.

---

[22] https://kantarainitiative.org/confluence/display/uma/Home

Similarly, Watson *et al.* [56] reflect on collective management of cybersecurity for a range of resources (devices, social media, streaming accounts etc). Whilst not specifically for smart homes, a key concern raised is those inside the group posing threats to group as a whole, in part due individual to poor security practices compromising the group as a whole. Whilst in practice, the participants hold each other accountable for shared resources, 'group security is only as strong as the individual with the weakest security behaviours, which can be inequitable and lead to resentment'. Thus, each member has responsibility, even if unsupervised by others to do their best, and they can risk losing access to these resources. In contrast to papers above flagging concerns around hierarchies, they suggest need for tools to address this weakest link element, and to allow 'group members with higher S&P awareness, motivation and knowledge to act as stewards for more vulnerable members.

Whilst this may be pragmatic, as we see above, vesting too much control in one party can lead to other risks, and impacts for the domestic social order. With DDCs navigating responsibilities, clearly there will be questions around to what extent collective management makes sense (e.g. in flat shares) vs one party managing security on behalf of everyone (e.g. with a family or elderly user). However, this should not just be because of the assumptions that a smart device driver will do this, or because of the affordances of a system around accounts.

Irrespective of if there is individual or collective control over devices in the home, one issue remains around the usability of systems. Adoption of technical solutions for accountability, such as the security systems mentioned above, need to be usable. In the enterprise context, there can be a perception that users are to blame for security not being managed effectively, but adopting user centric design principles or communicating how to use a system properly could address these [57]. Dourish *et al.* [58] examined system security manifests in everyday life for users. They focus on the disconnect between user goals and security goals can often leave users unsure how best to secure their systems. They state security for ubiquitous computing environments needs to be 'manageable and understandable to end users.

There are attempts to support user preferences around privacy practices for IoT with privacy assistants. Colnago *et al.* [59] qualitative study of user needs from privacy assistants showed different user preferences around degrees of automation, control and frequency of notifications. They advocate systems be built with modularity and configurability of autonomy for users to tweak systems individually. Part of this is assistants providing external recommendations, which is preferred to recommendations based on past behaviour, although the systems might need to provide choice over what recommended sources are used (again enabling tweaking by the user).

Nevertheless, this may change over time. Jakobi *et al.* [60] found the level of information users seek whilst managing smart homes differs over time. At the beginning there is a desire for more granular information and feedback on current and past behaviour. As years passed, this reduced to wanting information when systems 'not working, needed their attention, or required active maintenance'. How Article 5(2) requirements may require more long term oversight might not intersect with domestic practice. Thus, there is scope for the law to disrupt how users might ordinarily live with smart technologies. In attempting to subject the home to regulatory norms, like accountability, this may impact how systems are integrated into daily life by requiring mechanisms that prolong system oversight and may even disrupt the social order

*2.4.6  Differentiated responsibilities*

We have seen the narrowing of the household exemption and joint controllership arrangements emerging, leading DP law to enter the home.

DDCs managing data collection in the home are still reliant on data processing architectures defined by others. This is unlike the normal situation for a controller where they design and define the nature of processing. Clearly, vendors of IoT devices, or those providing security management tools have a role to play in shaping how DDCs manage data domestically. They may even be joint controllers, depending on if they process personal data too, e.g. via analytics, cloud storage etc.

In any case, if they are joint controllers, they too are implicated in the need to help demonstrate accountability with GDPR. As this section has shown, the provision of security management tools themselves could help to provide technical measures required. We conclude by thinking about how accountability obligations for security might be shared by vendors and DDCs.

There is a clear role for vendors in helping DDCs manage data processing in the home. Two key areas that have emerged are around improving quality of account management tools. Current approaches do not reflect the context of use and complexities of permissions, accounts and access. Another is around ensuring that there is usability of security tools. Whilst there are promising approaches emerging, increased engagement with DDCs and users is necessary to see the competencies, skillsets and challenges they face in using these systems. Some are technical, where users may lack those skills; others may want more control. Vendors should do more in integrating such systems into their offerings, including using user centric design principles to shape what this looks like. Such a supportive role results not just from the fact that vendors tend to be better resourced and skilled in managing security threats, but also more importantly, they are in a unique position to have native, usable tools built into their products to achieve and demonstrate compliance. There is also an economic case to be made here, considering how vendors may be held jointly liable for data breaches caused by security incidents [61]. Broadly speaking, one can even argue what accountability means for vendors is two-fold: To be able to demonstrate their own compliance, and to be able to demonstrate usable solutions provided to DDCs to facilitate their compliance.

With or without the support from vendors, there is also a clear role for DDCs in demonstrating accountability. On one hand, this may be as simple as adopting the use of technical measures, systems like Aretha, DADA, IoTSentinel, and relying on this as a demonstration of what they are doing to secure the home. However, as discussed, the home is a complex organisational space for these technologies to be used. They also have to integrate with how security of IoT devices is currently managed in the home. This includes thinking about interpersonal relationships, control, power asymmetries and guarding against abuses of such power. Ordinarily, such practices would not be mediated by legal requirements and further work will be necessary to see how DP law might restructure the social order of the home, by introducing new formally defined accountability norms.

## 2.5 Overall conclusions

In conclusion, within this paper, we began by exploring the nature of the accountability principle and thinking about security compliance obligations in GDPR. We then charted two fundamental shifts in recent EU DP case law: the narrowing of the household exemption in DP law and the broadening of notions of joint data controllership. The consequence of this is a new class of domestic data controller with responsibilities to other home occupants when processing their personal data. To contextualise this shift, we considered the implications for accountability in smart homes, and how home occupants who find themselves as DDCs might demonstrate compliance with security requirements of GDPR. We focused on the organisational setting of the home, as a site of smart technology use, and the types of technical tools available to DDCs that could provide a demonstration of accountability. However, we also raised complex questions that need to be addressed in order for DDCs to do this effectively. These ranged from unpacking the nature of interpersonal relations in the home around power, control and access; how technical tools might intersect with current security concerns and practices; and the differentiated nature of responsibility between joint controllers, the vendors of smart home technologies and the DDCs.